\documentstyle[psfig,referee]{mn}

\newif\ifAMStwofonts

\ifoldfss
  \ifCUPmtlplainloaded \else
    \NewTextAlphabet{textbfit} {cmbxti10} {}
    \NewTextAlphabet{textbfss} {cmssbx10} {}
    \NewMathAlphabet{mathbfit} {cmbxti10} {} 
    \NewMathAlphabet{mathbfss} {cmssbx10} {} 
  \fi
  \ifAMStwofonts
    \ifCUPmtlplainloaded \else
      \NewSymbolFont{upmath} {eurm10}
      \NewSymbolFont{AMSa} {msam10}
      \NewMathSymbol{\upi}     {0}{upmath}{19}
      \NewMathSymbol{\umu}     {0}{upmath}{16}
      \NewMathSymbol{\upartial}{0}{upmath}{40}
      \NewMathSymbol{\leqslant}{3}{AMSa}{36}
      \NewMathSymbol{\geqslant}{3}{AMSa}{3E}

      \let\leq=\leqslant 
       
    \fi
  \fi
\fi 

\ifnfssone
  \newmathalphabet{\mathit}
  \addtoversion{normal}{\mathit}{cmr}{m}{it}
  \addtoversion{bold}{\mathit}{cmr}{bx}{it}
  \newmathalphabet{\mathbfit} 
  \addtoversion{normal}{\mathbfit}{cmr}{bx}{it}
  \addtoversion{bold}{\mathbfit}{cmr}{bx}{it}
  \newmathalphabet{\mathbfss} 
  \addtoversion{normal}{\mathbfss}{cmss}{bx}{n}
  \addtoversion{bold}{\mathbfss}{cmss}{bx}{n}
  \ifAMStwofonts
    \ifCUPmtlplainloaded \else
      \UseAMStwoboldmath
      \makeatletter
      \new@mathgroup\upmath@group
      \define@mathgroup\mv@normal\upmath@group{eur}{m}{n}
      \define@mathgroup\mv@bold\upmath@group{eur}{b}{n}
      \edef\UPM{\hexnumber\upmath@group}
      \new@mathgroup\amsa@group
      \define@mathgroup\mv@normal\amsa@group{msa}{m}{n}
      \define@mathgroup\mv@bold\amsa@group{msa}{m}{n}
      \edef\AMSa{\hexnumber\amsa@group}
      \makeatother
      \mathchardef\upi="0\UPM19
      \mathchardef\umu="0\UPM16
      \mathchardef\upartial="0\UPM40
      \mathchardef\leqslant="3\AMSa36
      \mathchardef\geqslant="3\AMSa3E

      \let\leq=\leqslant 

    \fi
  \fi
\fi 

\ifnfsstwo
  \DeclareMathAlphabet{\mathbfit}{OT1}{cmr}{bx}{it}
  \SetMathAlphabet\mathbfit{bold}{OT1}{cmr}{bx}{it}
  \DeclareMathAlphabet{\mathbfss}{OT1}{cmss}{bx}{n}
  \SetMathAlphabet\mathbfss{bold}{OT1}{cmss}{bx}{n}
  \ifAMStwofonts
    \ifCUPmtlplainloaded \else
      \DeclareSymbolFont{UPM}{U}{eur}{m}{n}
      \SetSymbolFont{UPM}{bold}{U}{eur}{b}{n}
      \DeclareSymbolFont{AMSa}{U}{msa}{m}{n}
      \DeclareMathSymbol{\upi}{0}{UPM}{"19}
      \DeclareMathSymbol{\umu}{0}{UPM}{"16}
      \DeclareMathSymbol{\upartial}{0}{UPM}{"40}
      \DeclareMathSymbol{\leqslant}{3}{AMSa}{"36}
      \DeclareMathSymbol{\geqslant}{3}{AMSa}{"3E}

      \let\leq=\leqslant 

    \fi
  \fi
\fi 

\ifCUPmtlplainloaded \else
  \ifAMStwofonts \else 
    \def\upi{\pi}
    \def\umu{\mu}
    \def\upartial{\partial}
  \fi
\fi

\title[Chemical evolution: SPH cosmological simulations]
{Chemical evolution using SPH cosmological simulations. I:
implementation, tests and first results}

\author[M. B. Mosconi, P. B. Tissera,  D. G. Lambas and S. A. Cora]
{M. B. Mosconi$^{1}$, P. B. Tissera $^{2,3}$, D. G. Lambas $^{1,3}$,
S. A. Cora$^{3,4}$ \\
$^{1}$Grupo I.A.T.E., Observatorio Astron\'omico de C\'ordoba, Argentina.\\
$^{2}$ Instituto de Astronom\'{\i}a y F\'{\i}sica del Espacio, Casilla de Correos  67, Suc. 28, Buenos
Aires (1428), Argentina.\\
$^{3}$Consejo Nacional de Investigaciones Cient\'{\i}ficas y 
T\'ecnicas, Argentina.\\
$^{4}$ Observatorio Astron\'omico de La Plata, U.N.L.P., Argentina.}

\date{Accepted .....
      Received .....;
      in original form .....}

\pagerange{\pageref{firstpage}--\pageref{lastpage}}
\pubyear{2000}

\begin{document}
\maketitle

\label{firstpage}

\begin{abstract}

We develop a model to implement metal enrichment
in a cosmological context  based on the hydrodynamical
AP3MSPH code described  by Tissera, Lambas and Abadi (1997). 
The star formation model is based on the Schmidt law and
has been modified 
in order to  describe the transformation of gas into stars 
in more detail.
The enrichment of the interstellar medium due to
 supernovae I and II 
 explosions is taken into 
account
 by
assuming  a Salpeter Initial Mass Function 
 and different  nucleosynthesis models.
The  different chemical elements are mixed within the gaseous
medium according to the Smooth Particle Hydrodynamics technique.
 Gas particles can be enriched by 
different
 neighbouring
 particles at the same time.
We present tests of the code that  assess  the effects of
 resolution and  model parameters on the results.
We show that the main effect of low numerical resolution is to
produce a more effective mixing of elements, resulting in 
abundance relations with less dispersion.
We have performed cosmological simulations in a standard Cold Dark Matter
scenario  and we
present results of the analysis of the star formation and
chemical properties of
the interstellar medium and stellar population of the simulated
galactic objects. We
 show  that these systems
reproduce  
abundance ratios for primary and secondary elements
of the interstellar medium,  and the  correlation between
the (O/H) abundance and the gas fraction  of galaxies.
We find that star formation efficiency, the relative rate of 
supernovae II to supernovae I and life-time of binary systems as
well as the stellar nucleosynthesis model adopted affect
the chemical properties of baryons.
We have compared the results of the simulations with an implementation
of the one-zone Simple Model, finding significant differences in
the global metallicities of the stars and gas as well as their correlations
with dynamical parameters of the systems.
The numerical simulations performed provide a detailed description of the
 chemical properties of galactic objects formed in hierarchical clustering
scenarios  and proved to be useful tools to deepen  our understanding
of galaxy formation and evolution.

\end{abstract}


\section{Introduction}

In recent years, our knowledge of the high
redshift Universe has increased dramatically
allowing us to start constructing a picture of
how the different morphological types evolve
(e.g. Steidel and Hamilton 1992;
Madau 1995; Lilly et al. 1995; Cowie et al. 1996; Ellis et al. 1996;
Steidel et al. 1998).
In particular, the comoving star formation
history first depicted by Madau et al. (1996)
has resulted in a very useful way of
quantifying galaxy evolution, under certain hypothesis. 
Since star formation directly translates into
metal enrichment of the interstellar medium (ISM), observations of these two processes help
astronomers to constrain models of
the formation of the structure.
In particular, Lyman alpha forests, damped
Lyman alpha systems and Lyman alpha  break 
galaxies have all contributed to estimate
 the cosmic metal ejection and star formation  rates (e.g., Lu et al. 1996; 
 Pettini et al. 1997).
However, we are still far from drawing a consistent
picture since several points remain to be clarified. In this respect, the integration of
the dust corrected cosmic star formation history
of the Universe up to $z=0$ can account for
the entire stellar mass content in 
spirals and spheroids at present times. However, when
we look at the metal content at $z\simeq 2.5$,
only  $10\%$ of what is expected is actually
measured (Renzini 1998).
Regarding the chemical properties of galaxies, most of the available observations
are restricted to the Galaxy, and in particular, to the solar neighbourhood (e.g., 
Edvardsson et al. 1993; Gratton et al. 1996; Rocha-Pinto et al. 2000). Extragalactic
observations of HII regions  provide  information on 
the chemical content in other galaxies (e.g., Pagel 1992;
Garnett et al. 1995; Kennicutt and Garnett 1996; Kobulnicky and Skillman 1996).

Chemical evolution models are a useful tool to attempt 
to study the physical processes that might determine the chemical 
characteristics of the different galaxies. 
Detailed modelling
can be found in studies of the Galaxy (e.g., Ferrini et al. 1992; 
Tossi 1996; Chiappini, Matteucci and Gratton 1997 and references therein).
These analytical models carefully treat stellar evolution. However, they cannot
account for dynamical evolution, and certainly, not for hierarchical clustering.
From a numerical point of view, metallicity enrichment mechanisms have been implemented
in hydrodynamical simulations
in different ways. The first attempts were done 
by Larson (1975, 1976). Other implementations  came after using
diverse techniques such as
chemodynamical models (e.g., Burkert et al. 1992; Samland et al. 1997)
which describe in more detail the interstellar medium evolution.
On the other hand, this approach uses a very simple prescription for galaxy
formation  where the dark matter halo is either not considered at all or assumed to
be static. However, there are strong evidences from observations
of normal spirals that the dark matter is
dynamically important within the luminous radius 
 (e.g., Bottema 1992; Courteau, de Jong \& Broeils
  1996; Rhee 1996),
and that the response of the dark matter to the presence of 
baryons affect their evolution (Blumenthal et al. 1996; Tissera \&
Dom\'{\i}nguez-Tenreiro 1998) and, as a consequence, the star formation 
process (Mihos \& Hernquist 1996; Barnes \& Hernquist 1996; Tissera 2000;
Navarro \& Steinmetz 2000).
A first approach to include chemical evolution in Smooth 
Particle Hydrodynamical (SPH)
code is described by Steinmetz and M\"uller (1994, 1995)
followed by 
 Raiteri, Villata and Navarro (1996).
These  works  
run prepared-cosmological
 initial conditions where the formation and
evolution of only  one object was studied.
These models have shed light on some physical mechanisms that may control metallicity in galaxies, and
 have also shown that hierarchical clustering scenarios may form galactic objects that
resemble Milky Way-type galaxies from a chemical point of view.
However, the SPH models with chemical implementations hitherto published
 are restricted to the Galaxy. It would be very  important for the
study of the  formation and evolution of galaxies to be able to  analyse a variety  of  galactic objects
with different evolutionary histories.
Hydrodynamical cosmological simulations are more suited to 
tackle this problem since the non-linear evolution
of the matter is naturally accounted for, and,
physical processes can be more consistently implemented. 
These models can  provide coherent well-described 
environments for all objects and a complete record
of their formation and evolution. 
The drawback is the treatment of numerical resolution
effects. 

In this paper, we  concentrate on the description of the chemical model
that has been implemented in a cosmological hydrodynamical AP3M.
First results on galaxy formation, global chemical
properties and comparison with observations  are  reported.
 The detailed analysis of the 
chemical properties of the ISM and stellar population in galaxy-like objects
such as abundance gradients, age-metallicity relations, etc., 
are given by Tissera et al.  (2000, hereafter Paper II). 

This paper is organized as follows. Section 2 describes
the star formation process and the metal 
enrichment implementation. In Section 3 we assess the performance of 
the chemical  model. In 
Section 4  we present the first results on galaxy formation and chemical evolution. Section 5
summarizes the results.

\section{Numerical Model}

We use a cosmological numerical code based on the  
SPH technique
described by
Tissera et al. (1997). 
The SPH algorithm has been coupled to the AP3M gravitational code (Thomas
and Couchman 1992),
in order  to follow the gravitational and hydrodynamical evolution of particles 
within 
a cosmological context.
For the sake of simplicity, and  as  first step in the development of this chemical cosmological code, 
 simulations have been  run considering that
the gas cools down by using the approximation given by Dalgarno and McCray (1972).

\subsection{Star formation}


In general, the modelling of star formation (SF) in SPH simulations is based on the Schmidt law and a series of hypothesis
to select suitable  SF regions. 
However, how it is actually implemented within each particular  code depends
on the different authors (e.g., Katz 1992; Navarro and White 1994; Tissera et al. 1997).
 In this
paper, we describe a modified version of the SF implemented by Tissera et al. (1997) 
in order to be able to track the transformation of gas into stars in more detail, within a given
gas particle.

We include the SF algorithm as follows. Gas particles are eligible to form stars if they 
are cold ($T < T_{\star}$) and
satisfy a density criterium: $\rho_{\rm gas} > \rho_{\rm crit}$. This density criterium arises
by requiring the cooling time of a gas particle to be smaller than its dynamical time.
The critical temperature $T_{\star}$ is taken as the minimum provided
 value by the 
cooling functions ($T_{\star} \sim 10^4$K).
Finally,
gas particles have  to be part of a collapsing region. 
This requirement  is imposed  by selecting 
 particles in a convergent flow ($\bigtriangledown . \vec v < 0$).
When a gas particle satisfies all these conditions, star formation  occurs according to the Schmidt law,
\begin{equation}
 \frac{d\rho_{\rm star}}{dt}=c \,\,\frac{\rho_{\rm gas}}{t_{\star}},
\end{equation}
where $c$ is the
star formation efficiency
and $t_{\star}$ is a characteristic time-scale
assumed to be proportional to the dynamical time of the particle ($t_{\star} 
=t_{\rm dyn}=
(3\pi/16 G\rho_{\rm gas})^{1/2}$).
 Then, each stellar mass formed in a particle at a certain SF episode
is given by $\Delta_{\rm star}= C \,\,\rho_{\rm gas}^{3/2} \Delta t$,  
where $\Delta t$ is the time
step of integration ($\Delta t = 1.3 \times 10^{7}$ yr)
and $C$ a new SF efficiency.
Note that, according to equation (1), in order to go from density to  mass, 
the volume of the new-born stars has to be assumed. Taking into account
that the process we are modelling happens {\it within a gas particle} and, that  according to 
observations, SF occurs normally in clusters, not in isolation, we assume that
the volume occupied by the fraction of new-born stars is constant
 for all them.
Thus, it can be  absorbed in the  new constant, $C$.

According to this SF scheme, in a given baryonic particle there could be several SF episodes that had occurred
at different times.
The gaseous  mass of the baryonic  particle is reduced by 
$\Delta_{\rm star}$
 until
its  gas reservoir is depleted. A minimum gas mass equal to 5 $\%$ of the initial gas mass component
is left over in each particle.
 When a particle reaches this minimum mass, it  is transformed completely
into a star particle  and hereafter behaves  as collisionless matter.
Each $\Delta_{\rm star}$
formed can be followed up in time. 
The number of stars of a given mass within  $\Delta_{\rm star}$ is estimated
by assuming an Initial Mass Function (IMF). 
Hence, a baryonic particle may be formed by a gaseous and stellar components
in different proportions according to its history of evolution (hereafter, hybrid particle).
The stellar component can be  made up of different stellar 
populations with 
different ages and chemical properties, which have been formed at different SF episodes.
The chemical properties of each stellar population 
($\Delta_{\rm star}$) reflect the chemical state of the ISM 
at the time of its formation.

The decoupling between gas and stars depends on the actual  rate at which stars are 
formed in each baryonic particle, that, on its turn, depends on each evolutionary path.
Hybrid particles may introduce some numerical artifacts since the
stellar populations follow the gas evolution when they should probably not.
However, this is not a trivial problem that can be easily resolved.
An improvement of the decoupling process is under work and will be
presented in a separate paper.

\subsection{Metal Production}

Metals are produced and ejected to the interstellar
medium  at the end
of the life of stars. Most
of chemical  elements are ejected by Type II supernovae
(SNeII)
except for the iron that is mainly produced by
Type I supernovae (SNeI). 
%
Particles  are assumed to be
initially formed by Hydrogen and Helium
in primordial abundances (${\rm H=0.75}$, ${\rm He=0.25}$).
The first generations of stars with  primordial abundances immediately 
enrich the ISM from where the new generations are born.
It has to be stressed that we are not including the effects of (thermal
or kinetic) energy
injection into the ISM  due
to SN explosions. For the sake of simplicity we have
splitted the treatment of the feedback problem into two stages. Firstly,
we include chemical enrichment (this paper), and in a second step,
 we will intend
to develop an energy feedback model. There have been several
attempts to implement energy feedback in  SPH codes by either
injecting thermal  (e.g., Katz 1992) or kinetic (e.g., Navarro \& White 1994;
Metzler \& Evrard 1994; Navarro \& Steinmetz 2000; Springel 2000) energies to 
the gas component due to SN explosions.
However, these methods are still controversial leaving this problem
as an open question for galaxy formation.

Let us now describe the nucleosynthesis prescriptions adopted in this model:


\noindent {\it Type II SNe}:
We assume that stars more massive than $8 \ M_{\odot}$
end their lives as Type II SNe. 
In order to estimate their number, 
we adopt a Salpeter IMF with
a lower and upper mass  cut-off 
of $0.1$ and $120 \  M_{\odot}$,
respectively. 
The IMF adopted gives the 
total number of SNeII
formed in a certain range of stellar masses in a given $\Delta_{\rm star}$  at 
a certain 
time.
Supernovae are thought  to eject their
whole metal production within a few $ 10^7$ yr.
In particular, we assume that their life-time is equal to 
the integration time-step of the simulations.

For comparison, we resort to both  
Portinari, Chiosi and Bressan (1998, P98) and Woosley and Weaver (1995, WW95) metal ejecta models and
follow the evolution of different elements according to the information 
provided by the authors.
P98 give metal yields for stars up to 
$120 \ M_{\odot}$, while WW95  assume that stars 
larger than $40 \  M_{\odot}$ end up
their lives as black holes. We consider the following elements
 according to each author: 
 ${\rm H, He^4, C^{12}, O^{16}, Mg^{24},
 Si^{28}, Fe^{56}}$, for  both P98 and WW95, and also,  
$ {\rm N^{14}, Ne^{20}, S^{32}, Ca^{40}, Zn^{62} }$, for WW95.


{\it Type I SNe}:
Following Matteucci and Francois (1989), we assume
that supernovae Ia originate from carbon deflagration
in C-O white dwarfs in binary systems.
It is assumed that the masses of these binary systems
are
likely to be in the range $3-16 \  M_{\odot}$.
The adopted nucleosynthesis prescriptions 
are taken from Thielemann, Nomoto and Hashimoto (1993).
Type Ib SNe are assumed to be half the total
number of Type I SNe and to produce only iron 
($ \approx 0.3 \  {\rm M_{\odot}}$ per explosion).

The main
interest in including SNI events in the models
is that they contribute
with a substantial amount of iron.
This element is fundamental to attempt to   reproduce several observational results
(see also Chiappini et al. 1997).
The relative rate of SNI to SNII explosions 
has been observationally estimated in the solar neighbourhood 
and from extragalactic sources 
(van der Bergh 1991). 
These estimations suggest  a range of possible values for the 
total SNI rate respect to that of  SNII of  ${2 \leq\rm SNRII/SNRI \leq 3.11}$.

Binary systems evolve for a certain period of time ($t_{\rm SNI}$) 
during which mass is
transfered from the secondary to the primary star until the Chandrasekhar mass is exceeded and
a explosion is triggered.
It is generally assumed that   
a fair fraction of SNeI will explode after 
a period of  $t_{\rm SNI}\approx 10^8 - 10^9$ yr (Greggio 1996). 
It is straightforward
to estimate when SNeI will explode and
eject  metals to the neighbouring region, since 
the formation time of each $\Delta_{\rm star}$,
 in a given baryonic particle, is known.
The effects of varying both the SNRII/SNRI relative rate 
(hereafter, $\Theta_{\rm SN}$)
and $t_{\rm SNI}$ are analysed and results are confronted with
observations in the following Sections.

To sum up, the free parameters of our chemical model are:
  star formation efficiency ($C$), the initial mass function (IMF),
its upper and lower mass cut-offs,
  the  relative rate of different types of supernovae ($\Theta_{\rm SN}$),
  the life-time of binary systems that end as SNeI ($t_{\rm SNI}$) and 
 the yields.  Regarding the IMF, we are going to adopt a constant
and  unique one throughout this paper.

\subsection{Model for metal ejection}

How the chemical elements are distributed and mixed in 
the interstellar medium is a complex problem
which is far from being solved from a theoretical
point of view (e.g., Tenorio-Tagle
et al. 1999). 
Furthermore, numerical simulations
do not reach enough resolution to properly treat
this small-scale mixing process together with
the process of galaxy assembly. Hence, the ejection 
and mixing of elements have to be modelled in 
heuristic way.

We use the SPH technique to distribute
the metals within the neighbouring sphere of
the $j$  particle where a $\Delta_{\rm star}$ is
formed. According to this technique, the value of a certain parameter at the
location of $j$  particle   can 
be estimated by using the information storaged in its neighbourhood,
which is determined by
its $i$ neighbouring particles within its smoothing length ($h_j$).
   
Using this concept, we can define  the total mass ($M_j^k$)
 of a given $k$ chemical
element ejected by a $j$ particle   due to SNeI or SNeII,
as
\begin{equation}
M^{k}_{j}=\sum_{i=1}^{n_{\rm v}}m_i/\rho_i M^{k}_{j} W(r_i-r_j,h_{ij}),
\end{equation}
where $n_{\rm v}$ is the number of $i$ neighbours
 of the $j$ particle  ($n_{\rm v} \approx 40$), $m_i$ and $\rho_i$
the gas mass and density of the $i$ particle, respectively, 
$W(r_i-r_j,h_{ij})$ and  $h_{ij}$, the symmetrized kernel and   smoothing
length, respectively.
In this scheme, each $i$  neighbouring particle gets a
contribution of each metal  defined by
\begin{equation}
M_i^{\rm k}= m_i/\rho_i M^{\rm k}_{j} W(r_i-r_j,h_{ij}).
\end{equation}

The  $j$  particle where  the stellar mass has just formed is
 also included in this  process.
The masses of H
and He are proportionally decreased 
according to the metal mass received 
by each particle, so that its total mass is always conserved.

The distribution of metals by using the SPH technique allows to
spread out metals and enrich gas particles that may have not experienced
 SF but are
 close enough to other SF regions, working as an effective mixing mechanism.
Note that gas particles can be enriched by more than
one neighbour at each time-step.
It has to be mentioned that numerical resolution would affect the
mixing of metals since it also affects the determination of the neighbours.
In Section 3, we will discuss this point in more detail.
In the present implementation, metal mixing occurs only tided to the SF process,
in the sense that, only new-born  elements are distributed.  

One of the advantage of this implementation is that particles move
according to the equations of 
gravitation and hydrodynamics, leaving behind the closed-box hypothesis, among others.
Particles with different 
astrophysical and chemical
properties are mixed, particularly during violent events such as 
interaction and mergers, so that reproducing observations becomes a more challenging  task.

\section{Tests}

We carry out a series of simple tests to assess 
the effects of both,  numerical resolution and 
 the variation of the different parameters 
($C$, $\Theta_{\rm SN}$, 
$t_{\rm SNI}$) on the chemical properties of baryons. We also
intend to evaluate the dependence of the results on the two
adopted nucleosynthesis models, P98 and WW95. 
For this purpose, we follow the evolution of
a homogeneous  gaseous sphere. The sphere is
represented by $10^3$ gas particles initially distributed
following a density profile $\rho \propto r^{-2}$, in a radius of 625 kpc, 
without a dark matter halo.
This simple experiment gives a clear idea of how each parameter affects the 
chemical properties of
baryons and  have the advantage of a low computational cost.

\subsection{Effects of  numerical resolution}

Numerical problems may affect both, the process of star formation and 
specially the  distribution of metals. With respect to the former, if  dark matter haloes in  galactic objects
are  numerically well resolved,
the star formation process  may not be strongly
affected by the numerical resolution (see discussion in 
Dom\'{\i}nguez-Tenreiro, Tissera and  S\'aiz 1998). This is due to the fact that dark matter
determines the potential well on to which the gas settles on. If dark matter haloes 
are well resolved, baryons are forced  to distribute adequately. In this way, 
the gas density profiles are well reproduced and the   
 SF can be correctly followed (Tissera 2000). Nevertheless,
 the gas component has to be resolved, at least, by
few hundred particles (Navarro and White 1994).
In our simulations, baryonic and dark matter components are represented by 
particles of equal mass. Therefore,  dark matter haloes are resolved by a 
much larger number of particles than the baryonic component.  Hence, 
the potential wells of the larger galactic haloes are very well
defined ( see  Tissera and Dom\'{\i}nguez-Tenreiro 1998). 
Consequently, we restrict our global analysis to galactic objects resolved with more
than 250 baryonic particles.

On the other hand, the radius of the neighbouring sphere in which
the metals produced by the stellar component are
distributed is affected
by the numerical resolution of the experiment. Consequently,
this problem may have a non-negligible  effect on the metal distribution and
the chemical properties of gas and stars, and requires a more detailed
analysis. 
For this purpose, we have studied the evolution of two gaseous  
spheres without dark matter haloes. Each of them represents a system 
with total gaseous mass of $1.3 \times 10^{11} \ M_{\sun}$ and initial
radius of 615 kpc. One of the experiments (E1) resolves the sphere with
$10^3$ particles
(${\rm m_p} =1.3 \times 10^8 \ M_{\sun}$), while  
the second experiment (E1H) has  higher numerical resolution:
 $ 8 \times 10^3$ gas  particles (${\rm m_p}=1.7 \times 10^7 \ M_{\sun}$).
Since these spheres have no dark matter haloes, taking into account the
former discussion, it is reasonable to expect
that their star formation histories will be different.
 In fact, the sphere in E1H
starts forming stars before its counterpart in E1. 
However,  the effects that this earlier SF 
 has on their metallicity is not dramatic,
as it can be observed from Fig.1,
where we have plotted the  mean values of  [O/Fe] vs. [Fe/H] for both experiments. 
The greatest differences are obtained for the low metallicity stars 
([Fe/H] $< -3$) due to the fact that  star formation begins before in E1H and has 
a greater  intensity than  SF in E1.
As stars with higher metallicities are formed, this difference 
diminishes.
Note that  error bars in E1H comprise the distribution of E1 for 
[Fe/H] $> -4$ (except for the first point). 

The main difference observed between these two experiments is that 
  low numerical resolution  produces  
a  reduction in the dispersion of the distribution as can be
seen from the error bars in Fig.1.  
This fact can be understood by considering that, 
 in the experiment with lower resolution,
 the neighbouring  sphere 
($n_v \sim 40$)
comprises a larger volume, thus
producing a more effective mixing  of the metals in the gas.
Conversely, when numerical  resolution is increased, the new elements produced by a
certain particle are distributed in a smaller volume. This leads to
a less efficient mix at the scale of the system, producing  an increase
in the metallicity dispersion.
 We cannot theoretically assess at what extent this dispersion has any physical meaning since
the mixing mechanisms in galaxies are not yet well understood. It
might be possible that SN explosions spread metals in larger volumes
than that provided by the SPH technique. This fact would produce
a more uniform-metal distributions in the interstellar medium, as
it actually occurs in the low numerical resolution run. 
  

Taking into account these experiments, we 
conclude that the main effect of using the SPH technique to distribute metals
is that the 
 better resolved objects (i.e., those with higher number of baryonic particles that, in our simulations,
implies larger masses)
would have  a larger dispersion in their chemical properties  than the smaller ones. Mean values of
 the quantities will be considered as adequate  estimations.

\subsection{Effects of the variation of the star formation parameters}

We are now going to analyse the effects that the different
chemical model parameters have on the properties of the
stellar populations of the spheres.

As can be seen from Table 1, experiments  E1, E2 and E3 differ among each other in the life-time of 
binary systems that end up as SNeI. The values considered are 
$10^8$, $4 \times 10^8$ and $ 10^9$  yr, respectively. 
The SF efficiency and $\Theta_{\rm SN}$  used are the same in these three experiments. 
The ratio  [O/Fe] vs. [Fe/H]  clearly illustrates the effects
of varying the parameters. Fig.2 shows [O/Fe] vs. [Fe/H]       
for these three experiments at different stages of evolution.
As can be seen from this figure,  as soon as the first stars are formed, the differences among them are very 
important.
The smaller
$t_{\rm SNI}$ used in E1 allows a rapid production of Fe and, 
consequently, the number
of low metallicity stars ([Fe/H] $< -3$) is considerably lower than in
E2 and E3, for all  times.
The larger $t_{\rm SNI}$ value adopted in E3
produces a dex difference in the ratio [O/Fe] compared to the values in E1 and E2 for
$t=1.13 \times 10^{10}$yr.
As time goes on, all experiments reach solar values,
although, 
the variation of  
$t_{\rm SNI}$  among them leads
to appreciably different evolutionary paths. 
For E1 and E2, the [O/Fe] is above observed values 
for the solar neighbourhood, and the steepness of
the relation is much abrupt for 
[Fe/H]$ > -1$.
Experiment E1 ($t_{\rm SNI}= 10^8 $ yr) gives
a more consistent relation for stars with [O/Fe] $ <1$.
In this case, the final shape and range of abundances values are set
 as soon as the first
stars formed which implies that even at early stages of evolution,
this combination of SN parameters can produce stellar populations with metal
contents up to solar. 

In Fig.3, we show the same  relations for experiments E4, E5 and E6.
In E4   we
have changed the relative ratio $\Theta_{\rm SN}$ from
2 (in E1) to 4. Since SNRII is the fraction of stars with masses greater
than $8 \ M_{\odot}$ in a certain time-step, given by the IMF adopted, 
an increase in the ratio $\Theta_{\rm SN}$  implies a decrease of 
 SNRI. 
The effect of the variation of this parameter
gives the expected differences in the 
abundance tracks.
Note that there is no change in the slope of the [O/Fe] vs. [Fe/H] relation, 
 but the 
abundances in E4 do not get to solar
values because there are not enough SNeI to produce the necessary amount of Fe.

In experiment E5 we use a higher SF efficiency  than that in E1. 
In this case,  
stars are formed faster and
the gas is 
enriched very efficiently so that the new-born stars  immediately evolve  to
higher metallicity regions. As a consequence, 
the low-metallicity tail in the [O/Fe] vs. [Fe/H] relation
vanishes.
Note that the higher the $C$ value used, 
the larger the  stellar masses ($\Delta_{\rm star}$) formed, 
so that the number of SF events in a given baryonic particle
diminishes.
This is the reason  why the number of points in E5 is smaller 
 when compared to the distribution in E1.
The important aspect to remark is that, in E5,  the range of metallicity  
covered by the stellar populations is considerably smaller. 
This implies that a very efficient SF process can enrich the medium
to solar values very quickly so that most of stars would tend
to be located in the high metallicity region.

The comparison between experiments E1 and E6 
clearly illustrates how  the simulation results depend on 
the two yield models, P98 and WW95.
The features of the [O/Fe] vs. [Fe/H] relation in E6
can be appreciated
in the bottom raw of Fig.3.
The main characteristic observed for this experiment,
where  P98 yields were used, 
is the lack of the low-metallicity
tail ([Fe/H] $< -3$), which can be appreciated in E1. 
On the other hand, the low-metallicity stars in E6 ([Fe/H] $< -2$), have almost
constant values of [O/Fe], in contrast with the results of E1,
 where a gradient is observed.
This difference arises because P98 yields produce more iron with respect
to the  other elements than WW95, 
up to
an order of magnitude more, in some cases. 
So, stars are rapidly brought  to
lower abundances ratios with respect to iron. 
Despite this fact, these relations are quite similar, with  
exception  that 
WW95 seems to reach slightly higher metallicities values for the same 
model parameters at $z=0$. 
This fact implies that P98 yields for enriched stars 
 produce, in general,   more elements  
than those of WW95.
The comparison of other chemical 
elements shows that the major difference is in the 
[C/Fe] vs. [Fe/H] relation. 
In Fig.4, 
we show this relation for experiments E1  and E6  
as a function of 
time.
As clearly seen in the case of  E6, there is an excess of carbon in relation to iron for 
[Fe/H] $> -1.2$ that produces a positive gradient.

\begin{table*}
\centering
\begin{minipage}{140mm}
\caption{Test: Main Parameter
}
\begin{tabular}{@{}ccccc@{}}
S     &
$C$  &
$\Theta_{\rm SN}$    &  $t_{\rm SNI }$ &
yields\\ 
E1&   5e-6  & 2& 1 & WW95 \\
E2&   5e-6  & 2& 4& WW95\\
E3&   5e-6  & 2& 10& WW95 \\
E4&   5e-6  & 4& 1& WW95 \\
E5&   5e-2  & 2& 1& WW95 \\
E6&   5e-6  & 2& 1& P98 \\

\end{tabular}


Units: $[t_{\rm SNI}]=10^8 $ yr ; $[C]= Mpc^{9/2}/M_{\odot}^{1/2}/yr$.
\end{minipage}
\end{table*}

To sum up, we find that all SF and SN parameters  have non-negligible effects
on the chemical properties of baryons.
Then, in order to choose the correct combination of SN and SF parameters, a detailed
comparison with observations has  to be done.

\section{Galaxy Formation}

We are now interested in applying this model to the study of galaxy formation in 
a cosmological framework. 
As a first step, we look at the global properties of galaxy-like objects (GLOs) and compare
them with observations. The different formation and evolutionary histories of
each GLO (i.e., collapse time, merger tree, properties of progenitors, interactions, etc.) may
affect their SF and chemical content in a complex way so that, even  in the same
experiments, GLOs may exhibit different characteristics. This fact makes of
cosmological simulations a very useful tool for chemical evolution studies.

The traditional model to study chemical evolution is the so-called   One-Zone 
Simple Model based on
two main  assumptions: the system is isolated (i.e, no inflows or outflows) and 
well-mixed (i.e., instantaneous recycling), at all times (van der
Bergh 1962; Schmidt 1963; Tinsley 1980).
 Other  hypothesis are related to  the initial condition
of the gas, IMF and nucleosynthesis yields.
We remark the first two assumptions since they are inconsistent with the formation
and evolution of galaxies in a hierarchical clustering scenario. In our experiments,
 structures form  within a cosmological model suffering physical 
processes such as mergers, encounters,  inflows, etc.,  which may affect the dynamics
and kinematics of the dark and baryonic matters.
Although some authors have included inflows and relaxed the hypothesis of instantaneous
recycling (e.g., Chiappini et al. 1997), these models are not formulated 
in a cosmological framework and do not
include dynamical effects that may affect the SF process and the mixing of chemical
elements.
Regarding the IMF and  initial condition of the gas, we adopt a Salpeter
IMF for all times, and the gas is assumed to be 
initially in  primordial abundances.

\subsection{Numerical Experiments}

We performed SPH simulations consistent
with a Cold Dark Matter (CDM) spectrum with $\Omega =1$,
$\Lambda =0$,
$\Omega_{\rm b}= 0.1$, and
$\sigma_{8}=0.67$.
We used  $N = 262144$
particles ($ M_{\rm part}=2.6 \times 10^{8} \  M_{\odot}$) in a
comoving box of length $L=5 h^{-1} $ Mpc ($H_{0}=100 h^{-1}\ {\rm{ km \
s^{-1}\ Mpc ^{-1}}}$, $h=0.5$), starting at $z=11$.
Note that dark matter and baryonic particles have the same mass.
The gravitational  softening used in these simulations is 3 kpc, and
the smaller smoothing length allowed is 1.5 kpc.
Simulations  S1 to S5 share the same initial conditions while S6
 shares the SF and SN model parameters of S2, but has different random phases.

Simulations include SF and metallicity effects as described in  Section 2.
According to the discussion carried out in the previous Section, the value of
$C$, $\Theta_{\rm SN}$ and $t_{\rm SNI}$ affect the
chemical composition of the stellar and gaseous components.
In order to assess the impact of these parameters on  the chemical properties of galactic
objects, we have performed simulations with the same initial condition but varying the
model parameters (see Table 2). We also compare results from  simulations where 
 the two adopted yields, P98 and
WW95, have been used. 
We will focus on the study of global chemical properties of the ISM and
the stellar populations of GLOs, and their relation with the dynamical
parameters of the objects.

\begin{table*}
\centering
\begin{minipage}{140mm}
\caption{Cosmological Simulations}
\begin{tabular}{@{}ccccc@{}}
S &
$C$     &
$\Theta_{\rm SN}$  &
 $t_{\rm SNI } $ &
yields\\
S1&   5e-6  & 2& 1& P98 \\
S2&   5e-6  & 2& 1& WW95 \\
S3&   5e-4  & 2& 1& WW95 \\
S4&   5e-4  & 3& 5& WW95 \\
S5&5e-5 & 2 &1 & WW95 \\
S6&5e-6 & 2 &1 & WW95 \\
S7&5e-6 & 2 &1 & WW95 \\

\end{tabular}

Units: $[t_{\rm SNI}]=10^8 $ yr; $[C]= {\rm Mpc^{9/2}/M_{\odot}^{1/2}/yr}$. 
\end{minipage}
\end{table*}

In order to  study the properties of GLOs at $z=0$, 
we identify them at their virial radius ($\delta \rho/\rho \approx 200$; 
White and Frenk 1991).
In Table 3 we give their total dark matter  ($N_{\rm dark}$) and baryonic  ($N_{\rm bar}$) number particles
within the virial radius and their  virial circular
velocity ($V_{\rm vir}$). Letter b in the label code of the GLO (second
column, Table 3) indicates  if the main baryonic system is formed
by a pair of galactic objects. 
>From the set of GLOs identified, we are only going to analyse  those with more than
250 baryonic particles within their virial radius.

As it is well known, GLOs are formed by a dark matter halo that generally hosts
a main baryonic clump  and a series of satellites.
All quantities measured at the virial radius are related to these complex systems.
However, if we want to confront the simulated results with observations, it has to be
taken into account that observed astrophysical quantities come from the luminous matter.
For this purpose,  we define a galactic object (GAL) as the structure determined by
the main baryonic clump (including the dark matter mixed within it)
hosted by a GLO.

The radius that encloses $83 \%$ of the luminous mass of  an
exponential disc corresponds to the isophote of 25 mag ${\rm arcsec^{-2}}$.
Assuming the mass-to-luminosity ratio to be independent of radius, we define
the optical radius of a GAL 
as the one that encloses $83 \%$ of its baryonic mass. 
All chemical
and astrophysical properties will be referred to 
two optical radius ($2R_{\rm opt}$), 
unless otherwise stated (see Table 3).
This definition allows us to carry out a more meaningful comparison with observations. In this respect,
$V_{\rm opt}$,  defined as the circular velocity at $2R_{\rm opt}$,  
is determined by  the baryonic and dark  matter distributions in  GALs,
 and consequently, it is related to the physical mechanisms responsible
for the concentration and distribution of the matter in the central
region of a GLO (i.e., 
 star formation, mergers, gas inflows,  environment).
 Conversely,  $V_{\rm vir}$
is a global parameter determined by  the total potential well of the system.

Recall that, in these simulations, there are three types of baryonic particles: pure gaseous, hybrid and total
stellar ones. We identify those that belong to a given GLO and 
its GAL,  and look at
their chemical properties within $2R_{\rm opt}$ at $z=0$.

\begin{table*}
\centering
\begin{minipage}{140mm}
\caption{Galaxy-like Objects: Main Parameters
}
\begin{tabular}{@{}cccccccccc@{}}
S &
GLO &
$N_{\rm dark}$     &
$N_{\rm bar}$  &
$V_{\rm vir}$ &
$R_{\rm opt}$ &
$V_{\rm opt}$ &
$M_{\rm star}$&
$Z_{\rm gas}/Z_{\odot}$ &
$Z_{\rm star}/Z_{\odot}$\\
S1&  596& 4049 & 542 & 137.67 &  23.68 & 224.67 &   5.45  &   0.45  &   0.46\\
 & 538& 4574&  576 & 143.04 &  16.46 & 203.47  &  1.83    & 0.28  &   0.45\\
 & 538b& 4655 & 588 & 143.92&   19.16&  186.41 &   3.46 &    0.49 &    0.45\\
 & 536 &5036 & 636 & 147.74 &  19.89 & 243.13 &   6.25&     0.48  &   0.47\\
 &456 &6297 & 931 & 160.18 &   18.09 & 261.21 &   8.07 &    0.54 &    0.39\\
 &426 &2433 & 343 & 116.41 &  12.36 & 200.53 &   2.60 &    0.45     &0.44\\
 &421 &1848 & 267 & 106.11 &  19.61 & 166.06 &   1.43 &    0.26     &0.31\\
 &334 &1848 & 267 & 106.31 &  19.61 & 166.05 &   1.43 &    0.26     &0.31\\
 &347 &5752 & 724 & 154.41 &  14.06 & 246.20 &   5.46 &    0.61    & 0.45\\
 &347b& 5727 & 726 & 154.20 &  12.22 & 196.96 &   2.56 &    0.49   &  0.51\\
 &325 &5385 & 699 & 151.23 &   8.28 & 280.88 &   5.25 &    0.49  &   0.36\\
 &312 &1858 & 271 & 106.54 &  14.08 & 190.47 &   2.12 &    0.46 &    0.46\\
 &221 &1991 & 285 & 108.93 &   7.63 & 208.03 &   3.14 &    0.56&     0.47\\

S2&  596& 4034 & 541&  137.45 &  18.99 & 301.70 &   5.85  &   0.19  &   0.34  \\
&  538& 4591&  584&  143.29 &  18.74 & 251.69 &   2.37 &    0.17 &    0.24   \\
&  538b&4637 & 589&  143.77 &  11.25 & 246.80 &   3.24 &    0.21 &    0.27  \\
&  536& 5019&  627 & 147.46 &  14.82 & 319.25 &   6.57 &    0.20 &    0.30    \\
 &456& 6322&  932 & 160.35 &  21.63  & 323.83  &  7.36  &   0.19  &   0.23  \\ 
 & 426 & 1828 & 315 & 106.79 &   5.60&  290.85 &   4.60  &   0.37 &    0.36 \\   
 &421 & 2466 & 340 &  116.83 &  12.50 &  256.97 &   2.18 &    0.17  &  0.19 \\  
 &334 & 1863 &  266 & 106.48 &  23.28 & 203.43  &  1.67 &    0.09   &  0.24 \\   
& 347& 5770 & 720 & 154.52 &  14.64 & 308.19 &   5.41  &   0.27  &   0.29\\   
 &347b& 5715 & 721 & 154.10 &   9.19 & 249.24  &  2.73  &  0.22 &    0.28\\   
& 325 & 5478 & 703 & 152.01  & 12.91 & 315.70  &  5.37   &  0.23  &   0.21\\    
 &312 & 1840 &  271 & 106.24 &  15.67 & 239.76 &   2.23  &   0.19 &    0.21 \\  
 &221 & 1978 & 289 & 108.78 &  29.99 & 217.00  &  2.98  &   0.23   &  0.23 \\   


S3 & 596 & 4061 & 545 & 137.80 &  18.12 & 306.16 &   6.37 &    0.39  &   0.52 \\
  &538& 4570 &  590 & 143.16  &  9.67 & 258.72  &  3.42  &   0.59  &   0.64 \\
  &538b&  4541 & 588 & 142.85 &  22.15 & 246.53 &   2.68  &   0.20 &    0.34 \\
  &536& 5004 & 629 & 147.40 &  17.18 & 327.41  &  7.18   &  0.20  &   0.27 \\ 
& 456 & 6536 & 935 & 161.95 &  15.33 & 332.28 &  10.78  &   0.25  &   0.65\\
  &426& 2451 & 339 & 116.56  &  7.26 & 263.70  &  2.61  & 0.48    &   0.47\\ 
  &421& 1703 & 320 & 104.72  &  4.26 & 286.90  &  5.47  &   0.38  &   0.53 \\
& 334& 1847 & 273 & 106.00 &  18.54 & 210.09 &   2.23  &   0.17  &   0.60\\ 
 &347& 5874 & 724 & 155.37 &  38.15 & 337.76  &  9.57  &   0.16  &   0.29 \\
 &325 & 5502 & 714 & 152.32 &   9.28&  327.38 &   5.98  &   0.26  &   0.33 \\
& 312& 1876 & 275 & 106.89 &  13.97 & 239.87  &  2.63  &   0.17  &   0.26 \\
& 221& 2009 & 293 & 109.32  &  7.18 & 249.48 &   2.55  &   0.21  &   0.37\\ 
%
S4&  596& 4061 & 545 & 137.80 &  18.12 & 306.16  &  6.37  &   0.17  &   0.25\\  
&  538& 4570 & 590 & 143.16 &   9.67 & 258.72 &   3.42  &   0.18 &    0.25 \\
& 538b&  4541 & 588 & 142.85 &  22.15 & 246.53 &   2.68  &   0.13  &   0.25  \\
&  536& 5004 & 629 & 147.40 &  17.18 & 327.41  &  7.18  &   0.18 &    0.21 \\
& 456& 6536 & 935 & 161.95 &  15.33 & 332.28 &  10.78 &    0.21 &    0.32 \\
& 426 &2451 & 339 & 116.56&    7.26 &  263.70  &  2.61   &  0.16   &  0.19 \\ 
&  421& 1703 & 320 & 104.72  &  4.26 & 286.90  &  5.47   &  0.33   &  0.26 \\
& 334& 1847&  273 & 106.00  & 18.54 & 210.09 &   2.23   &  0.10  &   0.34  \\
& 347& 5874 & 724 & 155.37  & 38.15 & 337.76    &9.57  &   0.14 &    0.25 \\
& 325& 5502 & 714&  152.32  &  9.28 & 327.38 &   5.98   &  0.22 &    0.17 \\
& 312 &1876&  275 & 106.89 &  13.97 & 239.87 &   2.63 &    0.14 &    0.22 \\

S5&  596& 4036 & 538 & 137.51  & 22.34 & 297.28 &   6.49 &    0.19  &   0.37\\
&  538& 4629  &591&  143.66  & 17.70 & 241.27 &   3.45&     0.23  &   0.43\\
&  538b & 4520 & 577 & 142.55  & 720.66 & 247.62 &   2.43 &    0.14 &    0.26\\
&  536& 5037 & 640 & 147.75 &  14.53 & 328.11  &  7.38  &   0.24  &   0.33\\ 
& 456& 6326&  933 & 160.41 &  17.73 & 327.90  &  9.25  &   0.25  &   0.29 \\ 
& 426 &2446 & 344 & 116.59 &  11.46 & 259.62 &   2.85  &   0.24  &   0.25 \\
& 421 &1842 & 271 & 106.29 &  22.11 & 206.27 &   1.95  &   0.11  &   0.28 \\
& 334& 1842&  271&  106.29 &  22.11 & 206.27 &   1.95  &   0.11  &   0.28 \\
 &347& 5771&  723&  154.48 &  13.65&  309.75 &   6.35  &   0.30  &   0.37 \\
 & 347b&5722 & 727&  154.20 &   9.71 & 252.06  & 2.71  &   0.21  &   0.27\\
 &325& 5421 & 702 & 151.54  &  8.61 & 334.18  &  6.43&     0.27   &  0.26 \\

\end{tabular}
\end{minipage}

\end{table*}

\begin{table*}
\centering
\begin{minipage}{140mm}
\contcaption{}
\begin{tabular}{@{}cccccccccc@{}}
S &
GLO &
$N_{\rm dark}$     &
$N_{\rm bar}$  &
$V_{\rm vir}$ &
$R_{\rm opt}$ &
$V_{\rm opt}$ &
$M_{\rm star}$&
$Z_{\rm gas}/Z_{\odot}$ &
$Z_{\rm star}/Z_{\odot}$\\
S6&  665& 4552 & 626 & 143.30  &  7.54 & 287.46 &   3.44  &   0.20  &   0.21 \\
 & 565& 4408&  626 & 141.97 &   4.03 & 151.72   &  0.64   &  0.15   &  0.23\\
 & 546 &6452 & 895 & 161.04 &   6.14 & 357.70  &  6.86 &    0.22  &   0.19\\
 & 354&7421&  961 & 168.28  &  4.28 & 349.50  &  5.71   &  0.19  &   0.20 \\
 & 354b &2703 & 398 & 120.59 &  15.06 & 236.34  &  3.16  &   0.27  &   0.26\\
 &235& 3542 & 529  &132.27  & 11.10 & 310.54  &  6.03   &  0.34   &  0.27\\ 
 &234&1994  &275 & 108.84 &  17.48 & 212.84 &   1.94  &   0.16   &  0.21 \\
& 235b &4213 & 612 & 139.89 &  20.71 & 279.96 &   5.41  &   0.23   &  0.32 \\
& 127 & 6892&  797 & 163.51 &  21.66 & 329.06 &   7.08  &   0.21   &  0.28 \\ 
S7 & 554 &2748 & 385&  121.21&    8.98&  268.50 &   3.60 &   0.28 &   0.27\\
&  412 &2583 & 340 & 118.32 &   5.26 & 231.50   & 2.00  &  0.22 &   0.29\\
&  411 &2214 & 338 & 113.20 &   3.81 & 145.39   & 0.92  &  0.25  &  0.31\\
&  412b &4954 & 640 & 147.00 &  10.93 & 314.43   & 5.44  &  0.37  &  0.25\\
&344 &2673 & 317 & 119.35 &   5.93 & 282.17   & 4.44  &  0.49  &  0.34\\
& 261&2839 & 362 & 122.09  &  7.49 & 287.67   & 4.79  &  0.39  &  0.27\\
& 215 &3951 & 518 & 136.39  & 11.32 & 260.92   & 3.10  &  0.27  &  0.21\\
& 215b &3884 & 510 & 135.67  & 18.21 & 223.06   & 1.57  &  0.16  &  0.19\\
& 275& 10743 & 1417 & 190.48  & 14.20 & 382.86   & 13.14  &  0.27  &  0.28\\
& 232 & 8844 & 1204 & 178.73  &  5.95 & 403.29   & 7.78  &  0.16  &  0.18\\
& 235 &2162 & 244 & 110.95  &  3.50 & 195.64   & 1.93  &  0.42  &  0.25\\
& 233 &4002 & 574 & 137.52  & 10.37 & 257.12   & 3.73  &  0.30  &  0.31\\
& 245 &2206 & 258 & 111.88  &  8.03 & 250.97   & 2.97  &  0.30  & 0.30\\
& 222 &3148 & 412 & 126.46  &  8.00 & 286.05   & 5.11  &  0.36  &  0.30\\
& 225 &2474 & 370 & 117.37  & 13.47 & 258.48   & 3.98  &  0.26  &  0.22\\
& 131 &2292 & 244 & 112.97  & 18.75 & 191.21   & 0.57  &  0.09  &  0.11\\
& 131b &2335 & 244 & 113.59  & 31.22 & 220.19   & 1.52  &  0.17  &  0.21\\
& 134 &6934 & 981 & 165.10  & 11.23 & 365.57   & 8.53  &  0.22  &  0.25\\
& 135 &9700 &1311 & 184.27  & 10.60 & 364.38   & 6.89  &  0.17  &  0.19\\


\end{tabular}

$[V]={\rm km \ s^{-1}}$; 
$[M_{\rm star}]=10^{10}\  M_{\odot}$,
$[R_{\rm opt}]=$ kpc.

\end{minipage}
\end{table*}

\subsection{Global Chemical Properties}

In this Section we  study the correlations between chemical
and dynamical properties
of the GALs, in order to explore the
possible physical mechanisms
that may determine the metallicity of a galaxy,
and to assess the dependence on model parameters.

Observationally, defining the metallicity of 
a galaxy is a  complex matter since it depends on the
type and quality of available data, the element used as
the estimator, etc. (e.g., Kunth and Ostlin 1999).
Moreover, observations only give information on the metallicity of certain regions within a 
galaxy so that they could be considered 
good estimators of the global metallicity only if the ISM were efficiently
mixed. 

Let us first review the main hypotheses of the one-zone Simple Model
in order to understand the differences
and advantages of our chemical scheme.
The two basic assumptions of the Simple Model are: a) the system is isolated and  b) it is well mixed at all
times. It is also generally assumed instantaneous recycling. 

Concerning hypothesis a), our galactic objects form in consistency with a 
hierarchical clustering scenario 
via the aggregation of substructure. At $z=0$ a galactic object generally  consists of a dark matter halo, 
a main baryonic clump and a series of satellites. Even if the objects do not
 have nearby similar-mass objects,
they are rarely isolated. Moreover, 
there are always both, gas infall from the dark matter haloes
and encounters with 
 satellites that may tidally induce gas inflows as have been reported in
numerical (e.g., Mihos \& Hernquist 1996; Tissera 2000) and observational
works (Barton, Geller \& Kenyon 1999).
At higher $z$, interactions and mergers notably increase
so that the hypothesis of 'closed-box' is never valid since what
is observationally identified as a galaxy
is just a component of the whole system. It has also to be stressed 
that the star formation history
is affected by the evolution  of the galactic object. 
This evolutionary process  determines the
different chemical properties of 
the objects in our simulations 
(see Tissera et al. 2000 for details).

Due to the mixing mechanism adopted
which depends on the local gas density, and the  fact that the systems
are never isolated, the metallicity of the gas and the 
new-born stellar population is not uniform.
At a given time, we found a significant 
spread in the values of the ISM metallicites within  a given GAL
(Tissera et al. 2000).
These results are consistent with those 
from chemodynamical models (e.g., Samland et al. 1997).
Regarding instantaneous recycling, we follow the 
evolution of the populations according to the stellar masses
including supernovae I and II explosions. A consistent implementation
of SNeI is performed as described in Section 2.2. 
Planetary nebulae (PNe) have not been included in this work.

Unfortunately we cannot treat in as much detail the ISM as 
chemodynamical models do, but 
this drawback is compensated by the fact that galaxy formation 
is well-described according to a cosmological framework
without ad-hoc hypotheses.  
Moreover, the coupled non-linear evolution of dark matter and baryons 
has non-negligible effects
on the star formation 
(e.g., Navarro \& Steinmetz 2000; Tissera
et al. 2000) that is directly related to the enrichment process. 

We will first attempt to assess which dynamical parameters correlate with 
the chemical properties of the GALs. 
We define global quantities for the chemical content of GALs.
A  global  metallicity ($Z$) is assigned to the stellar population and to 
the gaseous component of a GAL,
 taking into account  all contributions
from either, the stars or the gas respectively, according to
\begin{equation}
Z_a= \frac{\sum_{k=1}^{n} M_k}{M_a},
\end{equation}
where $a$ refers to the component (i.e., stars or gas),
$M_k$ is the total mass of the $k$ chemical  element present in the $a$ component, $n$ is the total  number
of $k$ chemical elements considered, and $M_a$ is the 
total mass of the $a$ component within two  optical radius. The estimated $Z_{\rm star}$ and $Z_{\rm gas}$
for each GAL are listed on Table 3.

We have analysed the GALs formed in simulations S2, S6 and S7. These 
experiments have the same star formation and cosmological 
model parameters but different random phases in the initial conditions.
The SN parameters used in these experiments are taken from the best
results given by  the test runs ($t_{\rm SNI}=10^8$ yr, $\Theta_{\rm SN}=2$).
 In Fig.5a and Fig.5c we show  the global metallicities of the 
stellar population ($Z_{\rm star}$) and the gaseous component 
($Z_{\rm gas}$) of the GALs versus
their gas fractions. We find  the expected 
trend indicating that the smaller the left-over gas,
the higher the metallicity.
Note that the slope of this relation for the gas and the stars is
different, being steeper for the former.
 The averaged values  of global metallicities for these runs are:
$<Z_{\rm star}/Z_{\sun}>=0.25\pm 0.05$ and
$<Z_{\rm gas}/Z_{\sun}>=0.24\pm 0.09$. 
Note that we are considering the metal content of stars and gas particles
within $2R_{\rm opt}$ 
irrespectively of their particular location.
These mean metallicity values 
cannot be directly compared to 
that of disk stars in the solar neighbourhood. 

In Fig.5b and 5d, we plot $Z_{\rm star}$ and $Z_{\rm gas}$ versus
the total stellar mass $(M_{\rm star})$ within $2R_{\rm opt}$.
As can be seen there is no clear trend between these parameters, albeit
a  weak tendency to have the higher metallicities in both, gaseous and
stellar components of GALs with intermediate stellar masses: $3 \times 10^{10}
M_{\odot}$ to $6 \times 10^{10}M_{\odot}$.

For comparison, we have estimated these relations for our GALs assuming 
they have
behaved according to the Simple Model. We include three   cases:
$0.2Z_{\odot}$ (dotted-dashed lines),  $0.5Z_{\odot}$ (solid lines)
and $Z_{\odot}$ (dashed lines) \footnote{In Fig.5 (b,d), lines represent
the estimations of the Simple Model for a system with total mass
$M_{\rm bar}=10^{10}\ M_{\sun}$,  initially  with $M_{\rm star}=0$
and $Z=0$.} .
It can be appreciated the significant differences between the Simple Model
and the results of the numerical simulations. 
The slope  of $Z_{\rm star}/Z_{\odot}$
versus $M_{\rm gas}/M_{\rm bar}$ is different as well as the systematic increase
of stellar metallicity as a function of the stellar mass formed 
in the Simple Model which contrasts with the 
behaviour of the numerical simulations.
By inspection of the same relation for the gaseous components,
we find that  the Simple Model overestimates the metallicity of the 
galactic gaseous components for the parameters that match the chemical
distribution of the stars.
Hence, it seems that it is not possible to reproduce simultaneously 
the trends found in the stellar population and the gas of
the simulated GALs  with the Simple Model. This is somewhat expected
since our code takes into account more complex processes (such as 
mergers, interactions, gas infall, etc)  that are
not considered in the Simple Model.


We have  studied the dependence of $Z_{\rm star}$ on $V_{\rm vir}$ and
$V_{\rm opt}$ finding no significant correlations.
 We expected the lack of correlation with $V_{\rm vir}$ since it is
  determined by  the whole gravitational
bound system,
while the chemical properties of the 
GALs are more strongly related to the fate of the baryonic matter 
in the internal regions.
In fact, Fig.6a shows no significant trend between the stellar mass of the GALs and
their virial velocity. On the contrary, there is a clear correlation 
of $M_{\rm star}$ with $V_{\rm opt}$ as can be seen from Fig.6b. 
If a universal value for the stellar mass-to-light ratio is assumed, 
this correlation would imply a Tully-Fisher relation 
similar to those found in previous works (e.g., Tissera et al. 1997; Navarro \& 
Steinmetz 2000).
Given that our GALs follow 
the Tully-Fisher relation,  
one  would expect, in principle,   a correlation
of the stellar population metallicity with the $V_{\rm opt}$
which is not found.
We think that this lack of correlation is due to the fact that 
 the chemical properties of these GALs are also 
affected by their merger
history. Violent events have a strong impact on the
 baryon distributions and, 
consequently, on the metals mixing (White 1981; Cora et al. 2000).
Furthermore, these objects are continuously accreating gas  that 
contributes with pristine material to form new stars.
 From these results we find that the mean
metallicity of the stellar population of the GALs 
 cannot be directly linked to the
optical mass either.  
Note  the lack of SN wind effects in our simulations, as first discussed by
Dekel \& Silk (1986), could strongly affect the evolution of the gaseous
component of low virial mass haloes, producing a non-negligible effect
on the chemical properties of
these systems.

We have also studied how the properties discussed above change 
for different nucleosynthesis yield models.
We have analysed the relations shown in Fig.5  for
experiment S1 which has been run 
with the same SF and SN  parameters than those used in S2, 
but with the  nucleosynthesis model of P98 instead of those of WW95. 
It can be seen in Fig.7a,  a 
similar correlation than that shown in Fig.5a although
with a shallower slope; and   
a lack of correlation between $Z_{\rm star}$ and $M_{\rm star}$
in  Fig.7b.
It can also be appreciated that global metallicities in the simulation 
with P98 model are higher than those in simulations using
WW95 yields, with  average values of $<Z_{\rm gas}/Z_{\sun}>=0.40\pm0.11$
and  $<Z_{\rm star}/Z_{\sun}>=0.45\pm0.06$. 
Hence,  the different yields used imply significantly different results.

In order to explore the effects of changing
the  model parameters, $\Theta_{\rm SN},\  t_{\rm SNI}$ and $C$,
 we plot in Fig.8   
$Z_{\rm star}$ versus $  M_{\rm gas}/  M_{\rm bar}$  and $ M_{\rm star}$ for simulations  S3, 
S4 and S5 (see  Table 2).
By inspection to Fig.8a we can see that this correlation
is  present for all these experiments except S3.
This simulation has the  most efficient 
star formation, the largest value of $C$, which 
is an important parameter that strongly affect
star formation and metallicity:
for higher $C$ values, and keeping the
SN parameters fixed, the dispersion in metallicity 
increases. Note however  that if the rate of SNeI is
decreased (larger $\Theta_{\rm SN}$) and
the $t_{\rm SNI}$ increased, this correlation
is recovered (S4). Hence, these three parameters
are relevant at determining the global
chemical properties of the GALs. 
In the case of the relation between metallicity and luminosity (total stellar
mass), the correlation is  not present 
 as it can
be seen from Fig.8b. However, we
cannot further study  the relative importance of the model parameters
unless we look into more detail to 
the chemical properties of the gaseous and stellar
components (to be analysed in Tissera et al. 2000).

Estimates of the mean $Z_{\rm star}$ and $Z_{\rm gas}$ values of GALs
in each simulation (see Table 4) show that when the gaseous component
is  more gradually transformed into stars (S1, S2, S6, S7), the mean global
metallicity of stellar populations and the ISMs at $z=0$ are very
similar, independent of the nucleosynthesis yield models used.
Conversely, when the SF efficiency is increased (S3, S4, S5),
the difference between the mean $Z_{\rm star}$ and $Z_{\rm gas}$ becomes
very important. In this case,  most of the metals are locked into
stars, regardless of the SN parameters.
Hence, in our chemical model a very efficient SF process produces, on
average, ISMs considerably less metal-rich than the stellar populations
of the GALs.

However, it has to be noted that the mixing mechanism plays a key role
in determining this result. A different implementation  
could lead to a more efficient distribution
of metals in a shorter time-scale, quickly enriching the ISMs.
Nevertheless, the way chemical elements are mixed in the ISM is
still a controversial question that remains to be resolved
from a theoretical and observational point of view.

Unfortunately, we cannot directly compare these results with those derived
from analytical
or chemodynamical model since the latter focus to the study of the Galaxy 
and do not have a sample where dependences with the dynamical parameters of 
the galactic objects
could be studied. Raiteri et al. (1996), although describing a similar
chemical model as the one studied in this paper and using the  SPH technique,
do not consider
different galactic haloes so that we cannot  compare our findings with 
their results.

\begin{table*}
\centering
\begin{minipage}{140mm}
\caption{Mean $Z_{\rm star}$ and $Z_{\rm gas}$ values of GALs}
\begin{tabular}{@{}cccccccc@{}}
& S1&	S2 &	S3&	S4&	S5&	S6 & S7\\	
$<Z_{\rm gas}/Z_{\odot}>$&$ 0.45\pm0.11$&$0.21\pm0.06$& $0.29\pm0.14$&$0.18\pm0.06$&$0.21\pm0.06$&	$0.22\pm0.06$& $0.27\pm 0.10$\\
$<Z_{\rm star}/Z_{\odot}>$&$ 0.43\pm0.06$&$0.26\pm0.05$&	$0.44\pm0.15$&$0.25\pm0.05$&$0.31\pm0.06$&	$0.24\pm0.04$ & $0.26 \pm0.06$\\

\end{tabular}
\end{minipage}
\end{table*}

\subsection{Comparison with Observations}

Observations of galactic and extragalactic HII regions and OB associations 
provide
information about the chemical properties of the forming stellar population and
the ISM. These observations give relations for  some primary elements such
as S, Ne, and C, and  for the so-called secondary ones, like nitrogen, 
as a function of the ratio O/H.
In this Section, we resort to these observational  
results to assess the global chemical
properties of the stellar populations and ISM of the simulated GALs.

Concerning primary elements, Galactic and extragalactic
HII regions show that the ratios (S/O) and (Ne/O) do not depend on (O/H).
However, the dispersion in these ratios is quite large
(Pagel 1997). The Simple Model actually predicts a similar 
behavior since in this scheme, the ratios between primary elements are constant. 
However, this model fails to reproduce the observed correlation
between (C/O) vs. (O/H). 
A similar problem is detected
for the observed (N/O) vs. (O/H) that exhibits a steeper correlation 
than that predicted by this model.
Note that most observations are obtained from HII regions
which are assumed good tracers of the metallicity of the galactic ISM.

We analyse the ISMs in our GALs,
which are determined by the gas particle properties within $2R_{\rm opt}$.
The global 
metallicity,  $Z^k_{\rm gas}$, for  each $k$  chemical
element  in the ISM of a GAL 
is defined as follows,
\begin{equation}
Z^k_{\rm gas}=\frac{\sum_{i=1}^{\rm n_p} m^{k}_i}{M_{\rm gas}},
\end{equation}
where  ${\rm n_p}$ is the total  number of $i$ gas particles within
 $2R_{\rm opt}$,  $m^{k}_i$ is the mass of the $k^{th}$ element
in the $i^{th}$ particle and $M_{\rm gas}$ is the total
gas mass within $2R_{\rm opt}$. 
An equivalent relation can be defined for the stellar
population, $Z^k_{\rm star}$.

In Fig.9, we plot log(S/O) for the ISM in the   GALs  in 
simulations  S2, S3, S4
and S5.
 As it can be seen from this figure,  the 
mean values obtained in the simulations 
are in very good  agreement with observations
(taken from  Pagel 1997).
The average  log(S/O) for GALs shows  a large dispersion that compares
well with the observed values.
The larger S/O ratios are  obtained for the simulation with
the highest SF efficiency (simulation S3).   
However, by changing SN parameters in order to 
diminish the effects of SNI explosions,
and mantaining  the same SF parameter in the simulation (simulation S4),
we obtain substantially smaller abundance ratios and dispersions.

In Fig.10, we plot log(C/O) versus log(O/H) 
 for GALs in simulations S2, S3, S4 and S5 (WW95) 
and simulation S1 (P98).
We include observations of HII regions (Pagel 1997).
A remarkable fact that can be seen from this figure 
is  the large difference between the results of WW95 and P98 models.
GALs in S1, which are exactly equivalent to those in S2,
except for the nucleosynthesis  model
adopted, have ratios almost half an order of magnitude larger than their WW95 counterparts, and
are out of the observational range.
Conversely, GALs in any of the WW95 runs have  ISMs with abundances
 similar to observed ones. However, they
do not show the gradient  reported from observations of Galactic 
and extragalactic HII regions (e.g., Garnett et al. 1995;
 Kennicutt and Garnett 1996; Kobulnicky and Skillman 1996),
although the simulated abundances extend only in the range 
$-3.8 < $log(O/H) $< -3.2$. The simulated mean values are lower
than the observed ones but it has to be recalled that we are not including
PNe yields that are thought to be important contributors of C (and N).

Observations of (N/O) in HII regions show that this ratio 
increases with (O/H) (e.g., Pagel 1992).
 A gradient in the secondary
element is actually predicted by the Simple Model, although the steepness of the predicted
relation is much larger than that actually observed in HII regions. 
There are several hypothesis that may
explain the origin of these differences (Pagel 1997).
We estimate this relation for  GALs in WW95 runs. As shown in Fig.11, 
the simulated
ISMs have abundances that are in  good agreement with observations.
However, we only find GALs with   $12 +$ log(O/H) $<8.5$,   while
the observed gradient in  HII regions is  $12 +$ log(O/H) $ >8.5$.
We do not find GALs with average ISM metallicity corresponding to 
solar abundances, neither extremely low metallicity IZW18 type objects.
It should be considered, however, that 
only  massive GALs are analysed in the simulations.

Another important observational correlation is that between $12 +$ log(O/H) and 
 the gas fraction estimated 
for irregular and blue compact galaxies (e.g., Axon et al. 1998). 
This relation can be used to assess the metal content of the ISMs in relation
to the left-over gas mass of the corresponding GALs.
In Fig.12, we plot $12 +$ log(O/H) versus log $(M_{\rm bar} /M_{\rm gas})$ 
for the ISM in 
the simulated GALs ($M_{\rm bar}$ and  $M_{\rm gas}$ are the total baryonic and
gaseous mass within $2R_{\rm opt}$, respectively).
Observations taken from Pagel (1997) have been included for comparison.
The estimated abundance ratios in the models are consistent
with observations. It is notable that the  observed relation is well 
reproduced in the sense that objects with a lower gas fraction
tend to have larger metallicities without
the need to introduce other physical mechanisms 
such as SN energy injection effects to the ISM. 
The  Simple Model predicts a decrease of the metallicity with the gas
richness of the system but it fails to correctly reproduce the observational
relation.

\section{Conclusions}

We have implemented a chemical model in a SPH cosmological code. 
First results are reported together with
the assessment of the effects of  model parameters and  numerical resolution.
It is found that the major problem introduced by a low numerical 
resolution is the  artificial
smoothing of the chemical properties of the objects. However if a minimum number of baryonic particles
is imposed, then average values can be considered reliable.

The mean metallicity is found nearly independent of the total 
stellar mass of the system
indicating that the effects of 
mergers, interactions and gas infall 
on the mass distribution and metal mixing are very significant.
We find correlations between the stellar mass $M_{\rm star}$ and the virial 
and optical circular velocities  ($V_{\rm vir}$, $V_{\rm opt}$). 
However, the correlation
with $V_{\rm opt}$ is significantly tighter, consistent with
the Tully-Fisher relation. 
Nevertheless, the global metallicity of GALs, $Z_{\rm star}$  
shows no dependence on $V_{\rm opt}$, in spite of the fact that these
GALs satisfy 
 the Tully-Fisher relation and shows a strong
correlation between  the global metallicity of their stars and  gas  
components with the 
left-over gas fraction of the systems.
We only find a significant correlation between the global GALs 
metallicity of both
stars and gas with the left-over gas fraction of the systems.

An analysis of the GALs assuming they have
behaved according to the one-zone Simple Model
show important differences with 
the results of the numerical simulations. 
The slope  of $Z_{\rm star}/Z_{\odot}$
versus $M_{\rm gas}/M_{\rm bar}$ and the systematic increase
of stellar metallicity as a function of the stellar mass formed 
in the Simple Model contrasts with the 
behaviour of the numerical simulations.
We find that the Simple Model overestimates the metallicity of the 
galactic gaseous components if it is constrained to match  
the stellar chemical content.
According to the Simple Model, the global metallicity of a system
continuously increases with the stellar mass in disagreement with
the results found for the simulated GALs that show
higher metallicities for intermediate mass objects. 

We find that the star formation efficiency, SN model parameters and
the nucleosynthesis yields significantly 
affect the chemical properties of the GALs.
In those models where the SF process is gradual the mean stellar and gaseous
metallicities are similar. 
 
The observed abundance ratios for primary and secondary elements in
Galactic and extragalactic  HII regions 
 are naturally obtained.
However, the  ranges of the simulated global metallicities are smaller
than the observed ones.
Also, the  relation  between the (O/H) and  gas fraction obtained 
in the numerical simulations is consistent  with the observational results.

The suitable agreement between the models and the observations for 
these relations suggests
that  hierarchical clustering scenarios are able 
to reproduce the chemical properties of galaxies. 
This is a very encouraging fact taken into account that 
galactic objects formed by the accretion and mergers of substructures. 
Violent events are common
and ubiquitous, and can affect the dynamical evolution of the matter,
regulating the SF and chemical evolution in galactic objects.

We intend to improve this chemical model in the future by
allowing the gaseous component to cool according to its metallicity,  
and a further study of the mixing process
of chemical element in the ISM.
It is also under study a more efficient decoupling mechanism
between stars and gas that could shorten the period of hybrid state of a
baryonic particle. Finally, energy feedback remains  an open question
that we hope to address in a future work.


\section*{Acknowledgments}

We thank S. Woosley for kindly providing the yield tables and L. Portinari for 
clarifying some aspects of their nucleosynthesis model.
We acknowledge the careful reading and useful suggestions of the
anonymous referee.
We also thank   M. Abadi for useful discussions. P.Tissera
is  grateful to C. Chiappini for introductory discussion in chemical
evolution during the 1999 Aspen Summer Workshop.
The authors thank the hospitality of the Observatorio Astron\'omico
Centroamericano de Suyapa while finishing the writing of this paper.
 P.Tissera is grateful to the Observatorio 
Astron\'omico de C\'ordoba and the research group IATE
for allowing the use of their computational facilities.
M. Mosconi thanks IAFE for their hospitality during the preparation of this work. 
This work has been partially supported by CONICET, CONICOR,  SECYT and Fundacion Antorchas.

\clearpage
\section*{Figure Captions}

\indent Figure 1: Comparison of the [O/Fe] versus [Fe/H] relation obtained from the 
low (solid line) and high resolution tests (dashed lines).

\indent Figure 2:  [O/Fe] vs. [Fe/H] for experiments E1, E2 and E3 where the life-time of binary systems
has been varied: $10^8$ yr, $4\times 10^8$ yr and $10^9$ yr, respectively.
Time $t$ is given in units of $10^{10}$ yr.

\indent Figure 3: [O/Fe] vs. [Fe/H] for experiments E4, E5 and E6 where
the relative ratio of SNII/SNI ($\Theta_{\rm SN}$), the 
star formation efficiency   ($C$)   and the nucleosynthesis
yields   have been changed with respect to E1.
Time $t$ is given in units of $10^{10}$ yr.

\indent Figure 4: Comparison of the [C/Fe] vs. [Fe/H] relation for experiments E1: WW95  and
E6: P98, yields.
Time $t$ is given in units of $10^{10}$ yr.

\indent Figure 5: Global metallicities of the stellar population and
gas component  in GALs
as a function of  their gas fraction, $M_{\rm gas}/ M_{\rm bar}$, 
(a,c),
 and their total stellar mass ,$M_{\rm star}$, (b,d)  in simulations S2 (open pentagons),
S6 (filled squares) and S7 (open squares). Lines represent
the relations given by the Simple Model for $0.2Z_{\odot}$ (dotted-dashed
lines), $0.5Z_{\odot}$ (solid lines) and $Z_{\odot}$ (dashed lines).
 $M_{\rm star}$ is given in units of $10^{10} \  M_{\odot}$.

\indent Figure 6:  The total stellar mass of  GALs ($M_{\rm star}$)
as a function of  a) the virial velocity ($V_{\rm vir}$) and 
b) the optical velocity ($V_{\rm opt}$)
  for the same GALs shown in Fig.5.
 Velocities are  given in km${\rm s^{-1}}$ and 
$M_{\rm star}$ is given in units of $10^{10} \  M_{\odot}$.

\indent Figure 7:  Global metallicities of the stellar populations in GALs ($Z_{\rm star}/Z_{\odot}$)
in simulation S1 (P98)
as a function of  a) the gas fraction ($M_{\rm gas}/ M_{\rm bar}$),
and b) the total stellar mass ($M_{\rm star}$) of GALs. $M_{\rm star}$ is given in units of $10^{10} \ M_{\odot}$. Lines represent the relations given
by the Simple Model as in Fig.5. 

\indent Figure 8: Global metallicities of the stellar populations ($Z_{\rm star}/Z_{\odot}$)
as a function of  a) the gas fraction ($M_{\rm gas}/ M_{\rm bar}$),
and b) the total stellar mass ($M_{\rm star}$) of GALs
in simulations S3 (filled pentagons), S4 (open triangles) and S5 (open stars).
 $M_{\rm star}$ is given in units of $10^{10} \ M_{\odot}$.

\indent Figure 9:  (S/O)   versus  (O/H)  for the ISM of GALs in simulations S2, S3, S4 and
S5 (see Fig.5 and 8 for feature code).
 Observations of HII regions  
have been included (small filled circles).

\indent Figure 10: (C/O)  versus  (O/H)  for the ISM of GALs in simulations S2, S3, S4 and
S5 (see Fig.5 and 8 for feature code). GALs in S1 (filled triangles) have been included.
 Observations of HII regions  
have been included (small filled circles).

\indent Figure 11:  (N/O) versus (O/H) for the ISM of the  GALs shown in Fig.9.
 Observations of HII regions  
have been included (small filled circles).

\indent Figure 12:   Logarithm of the oxygen abundances  versus
the logarithm of their gas fraction for the same GALs shown in Fig.9.
 Observations of HII regions in   blue compact and 
irregular galaxies from Pagel (1997) have been included (small filled circles).

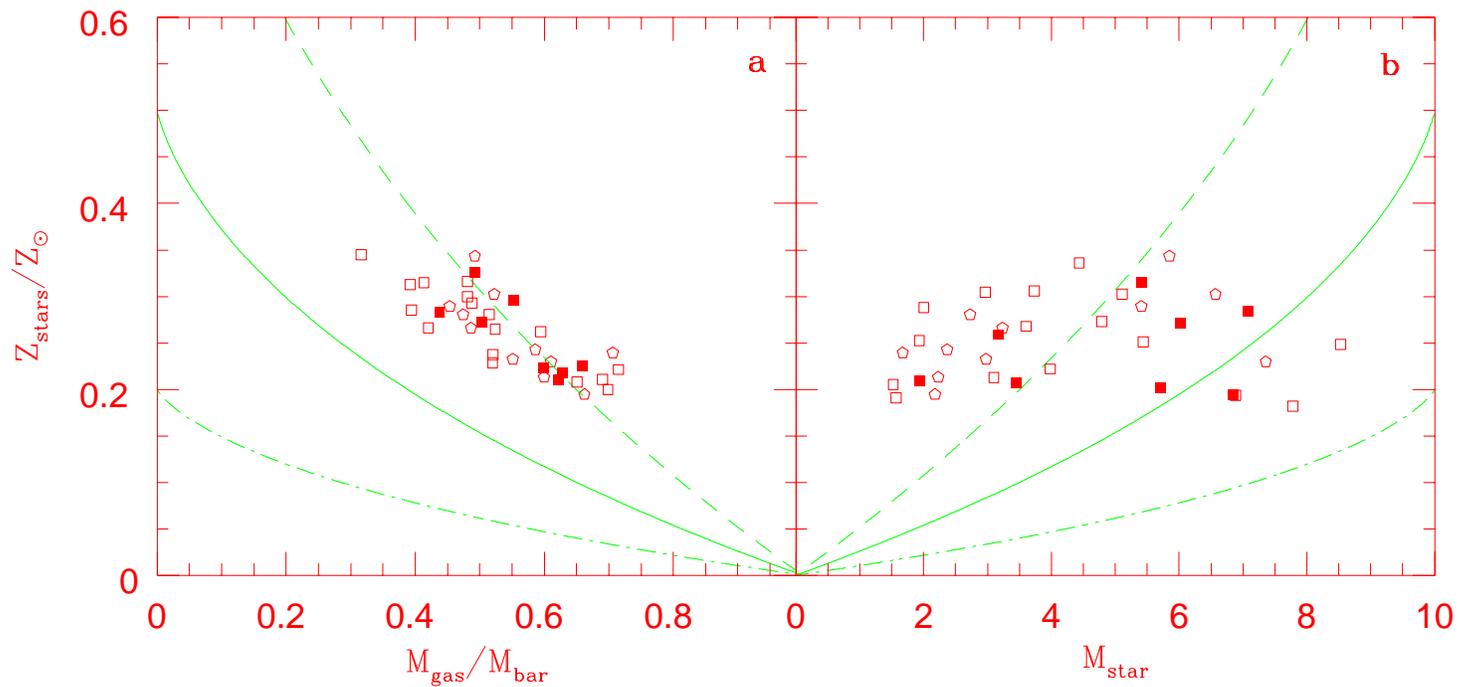
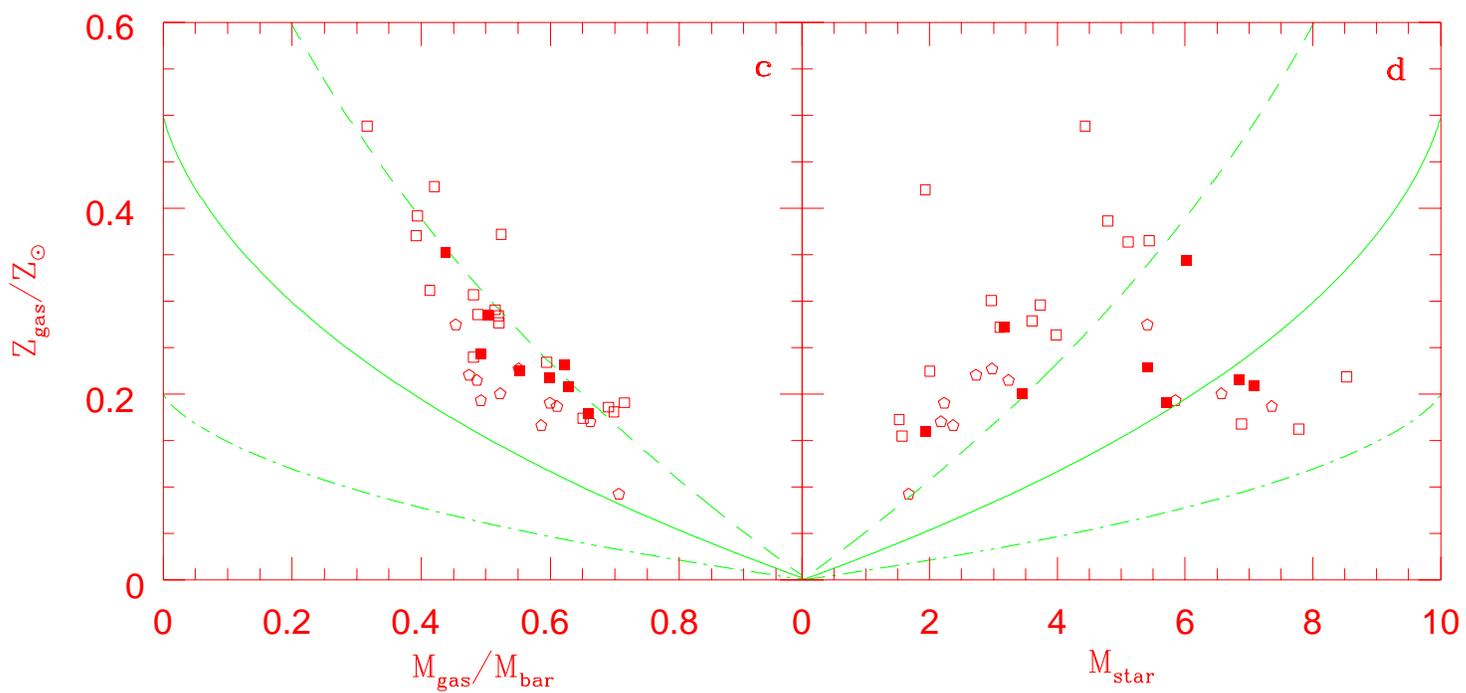
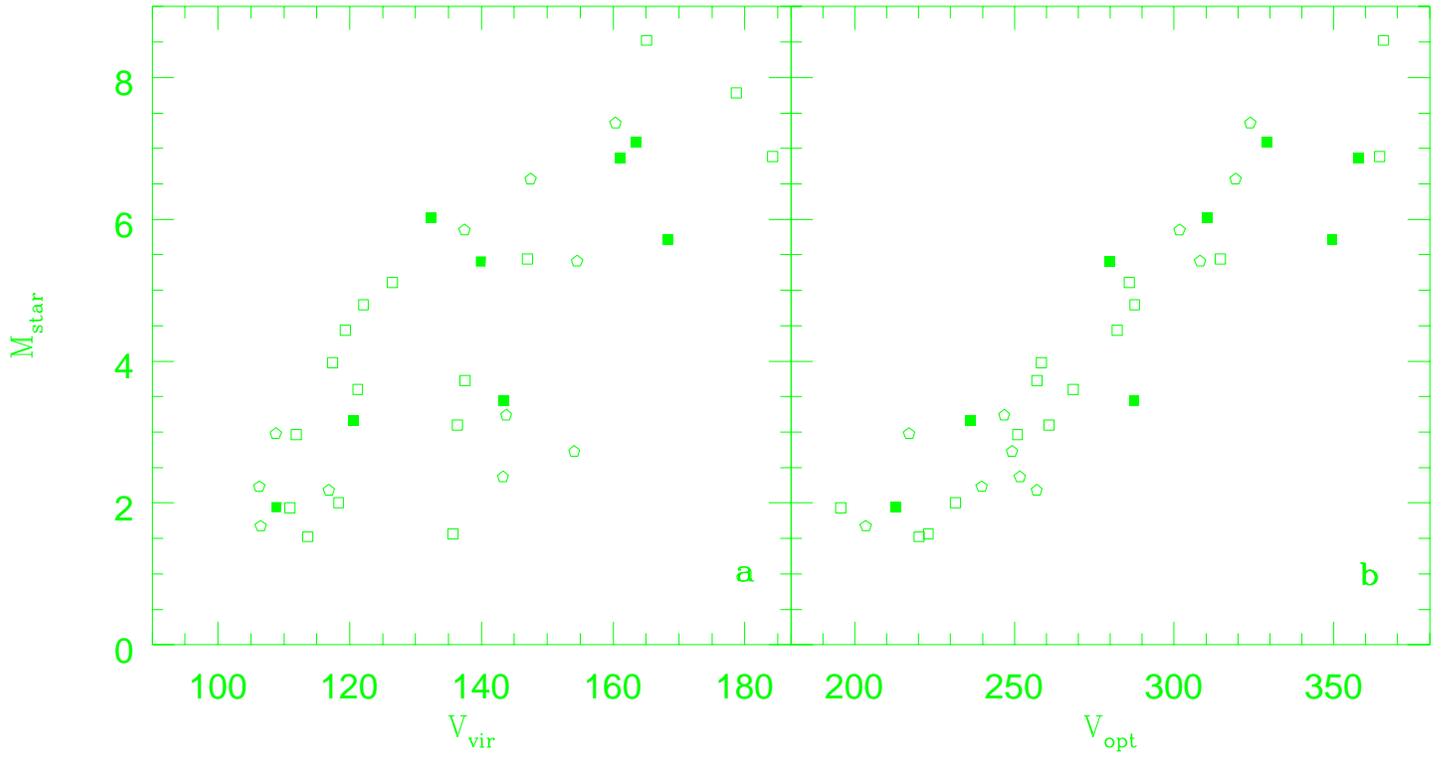
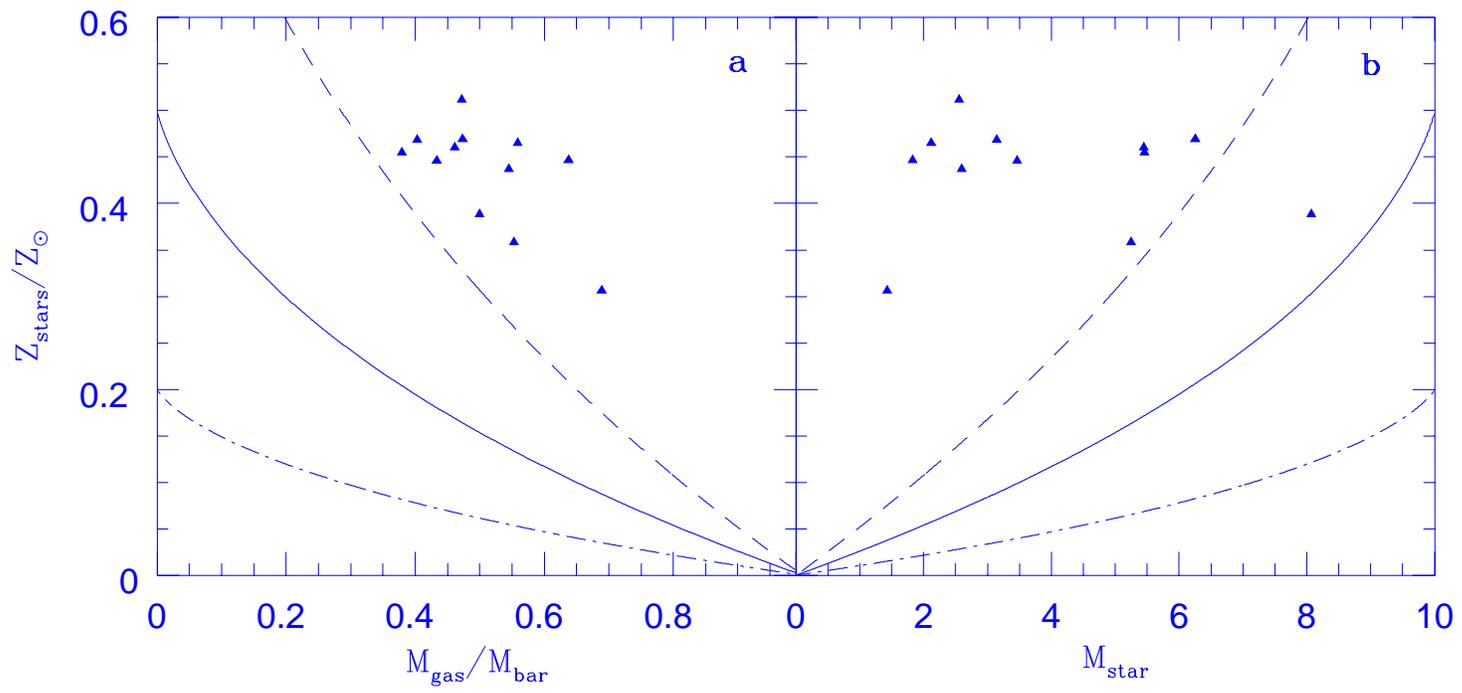
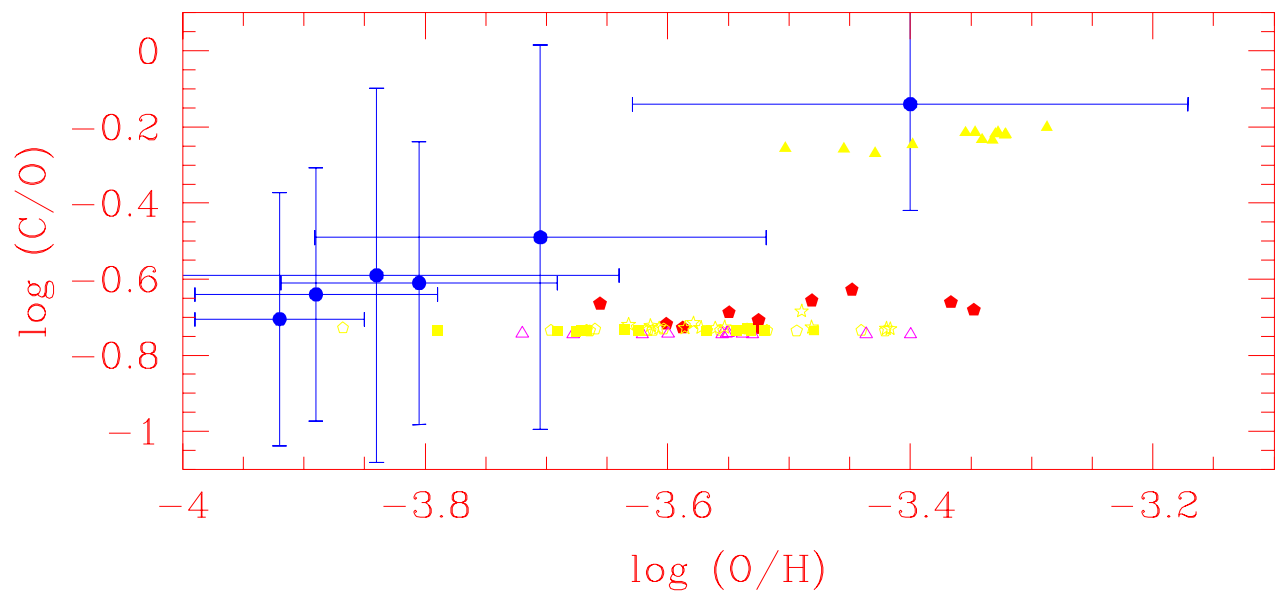
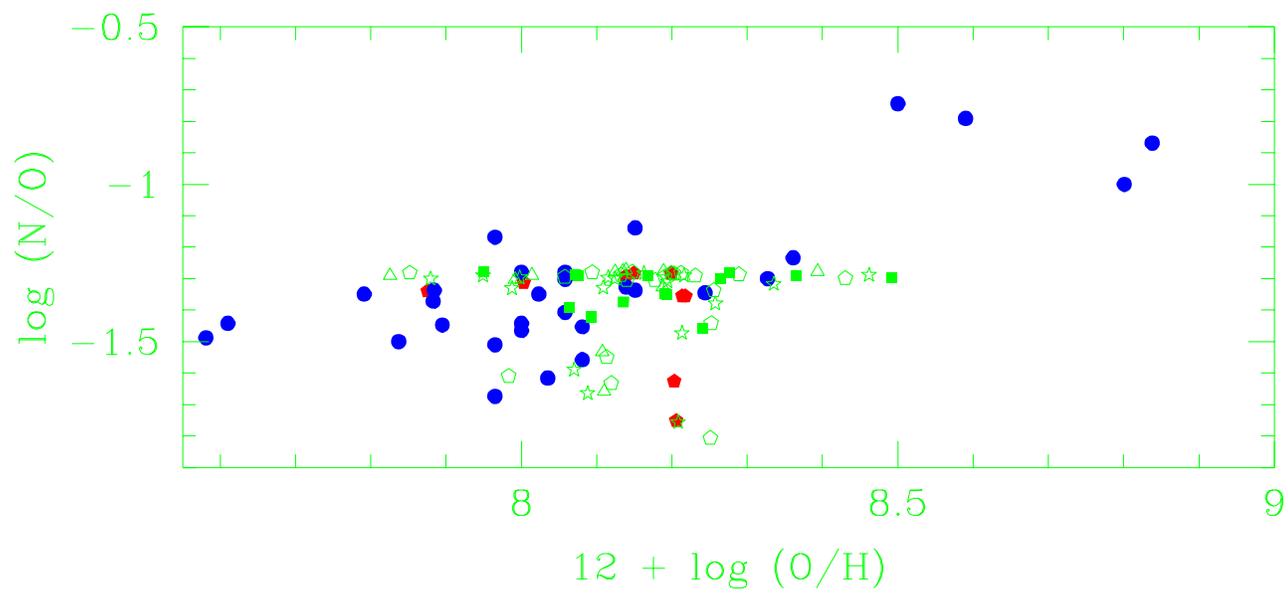
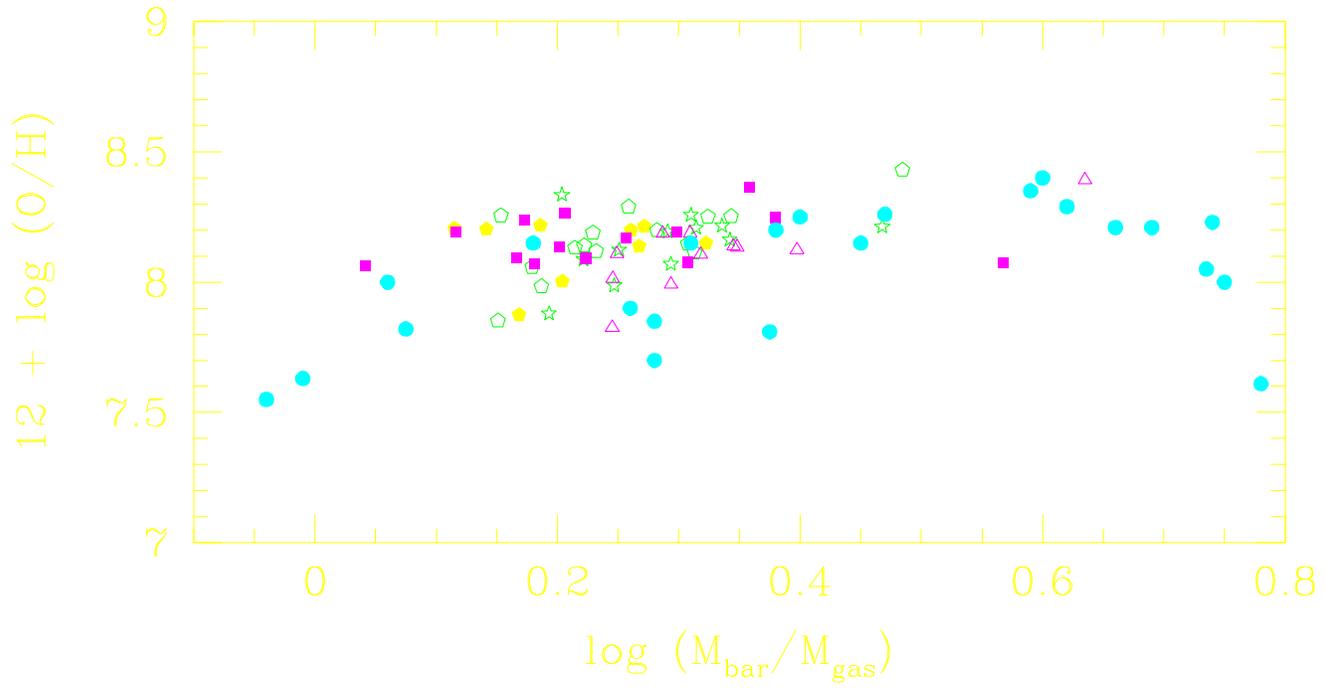
\end{document}